\documentclass[letter,twocolumn]{jpsj3}

\usepackage{txfonts}
\usepackage{amsmath,amssymb} 
\usepackage{graphicx,bm,color,comment}
\usepackage{fancybox}
\usepackage{empheq}
\usepackage{ulem}
\usepackage{mathrsfs}

\newcommand{\DIM}[1]{$#1 \msm1 \times \msm1 #1$}
\newcommand{\Equiv}{\msm2 \equiv \msm2}
\newcommand{\Eq}{\msm3 = \msm3}
\newcommand{\FRAC}[2]{{\textstyle \frac{#1}{#2}}}
\newcommand{\LT}{\msm2 < \msm2}
\newcommand{\MINUS}{\msm2 - \msm2}
\newcommand{\PLUS}{\msm2 + \msm2}
\newcommand{\bDD}{\bm{D}}
\newcommand{\bRR}{\bm{R}}
\newcommand{\baa}{\bm{a}}
\newcommand{\bdd}{\bm{d}}
\newcommand{\bkk}{\bm{k}}
\newcommand{\brr}{\bm{r}}
\newcommand{\bsfv}{\bar{\mathsf{v}}}
\newcommand{\bss}{\bm{s}}
\newcommand{\btt}{\bm{t}}
\newcommand{\dprime}{^{\prime \msm2 \prime}}
\newcommand{\hsfv}{\hat{\mathsf{v}}}
\newcommand{\hs}[1]{\hspace{#1}}
\newcommand{\msm}[1]{\mspace{-#1mu}}
\newcommand{\msp}[1]{\mspace{#1mu}}
\newcommand{\sV}{\mathscr{V}}
\newcommand{\sfC}{\mathsf{C}}
\newcommand{\sfF}{\mathsf{F}}
\newcommand{\sfX}{\mathsf{X}}
\newcommand{\sfm}{\mathsf{m}}
\newcommand{\sfv}{\mathsf{v}}
\newcommand{\tbee}{\tilde{\bm{e}}}
\newcommand{\ulCCC}{\underline{C}_{3}}
\newcommand{\ulV}{\underline{V}}
\newcommand{\ulone}{\underline{1}}

\title{Theory of Energy Dispersion of Chiral Phonons} 

\author{Hirokazu Tsunetsugu$^1$ and Hiroaki Kusunose$^{1,2}$}
\inst{$^1$The Institute for Solid State Physics, 
The University of Tokyo, 
Kashiwanoha 5-1-5, Chiba 277-8581, Japan \\
$^2$Department of Physics, Meiji University, Kawasaki 214-8571, Japan} 

\abst{
We have developed a microscopic theory on phonon energy dispersion
in chiral crystals within a harmonic approximation.
One of the main issues is about the splitting of sound velocity 
of acoustic phonons with opposite ``crystal'' angular momenta. 
We have shown that the splitting must be zero even in chiral 
crystals and the difference starts from the order of 
at least $k^2$ or higher in their energy dispersion.
Splitting is evident for chiral optical phonons, and we have 
derived a formula for their $k$-linear splitting.
Another important finding is about possible interactions of 
atomic displacements in microscopic models.
We have found that antisymmetric interactions of 
$\bDD _{ij} \cdot (\bdd _i \times \bdd _j) $ type 
are not allowed in microscopic Hamiltonians 
for chiral phonons 
because of the stability of equilibrium structure. 
We have identified that the splitting in both acoustic 
and optical modes arises from the harmonic potentials 
with the electric toroidal quadrupole of $G_{u}$-type symmetry.
These constraint are important for modeling real materials.
Most of our microscopic calculations have been performed 
for (quasi-)one-dimensional systems with a trigonal crystal 
symmetry including Te, but these results generally hold 
also for other chiral phonon systems.
}

\begin{document}
\maketitle

Chirality is a three-dimensional geometric concept 
that is defined by the absence of any mirror and inversion operations 
in systems under consideration\cite{Kelvin1904,Wagnie07}.
Dynamical aspects of chirality in materials 
and fields have been also emphasized by Barron\cite{Barron04}, 
and the microscopic definition of chirality has been recently 
introduced in terms of electronic multipoles\cite{Oiwa22,Kishine22}.
Since the proper rotation operations (and translations in crystals) 
alone cannot distinguish the difference between polar 
and axial properties, it allows systems to have a variety 
of couplings among axial and polar quantities 
such as electric field and angular momentum.
It leads to chirality specific cross-correlated responses 
discussed in many research fields such as biochemistry\cite{Hendry10}, 
nano-optics\cite{Gonokami05,Kelly20}, 
nonmagnetic inorganic 
crystals\cite{Rikken02,Tokura18,Furukawa18,Yoda18,Oiwa22}, 
and magnetism\cite{Muehlbauer09,Togawa12,Kishine15,Togawa16}.
Furthermore, the spin degrees of freedom 
even in nonmagnetic materials are
also involved in the so-called 
``Chirality-Induced Spin Selectivity'' (CISS), 
which has been actively studied
\cite{Gohler11,Naaman12,Naaman19,Naaman20,Evers22,Inui20,Nabei20,Shiota21}.

In particular,
phonons in chiral 
crystals\cite{Boer96,Ghosh08,Pine70,Pine71,Teuchert74,Martin76,Zhu18,Chen22} 
have attracted much interest 
because their transverse modes are characterized 
by the quantum numbers of 
proper ``crystal'' angular momentum (CAM)
about the chiral axis 
\cite{Vonsovskii62,Zhang14,Tatsumi18,Zhang22,Komiyama22,CAMcomment}.
The CAM related phenomena have been widely reported 
such as spin relaxation\cite{Garanin15,Nakane18,Streib18}, 
phonon-magnon conversion\cite{Holanda18,Guerreiro15}, 
orbital magnetization due to phonons\cite{Juraschek19}, 
phonon induced current\cite{Yao22}, 
and correction to the Einstein-de Haas effect\cite{Zhang14}.
The selection rule of CAM has been examined 
by the Raman scattering experiment\cite{Ishito21}.
Its relation to CISS has been also investigated 
as a source of spin filtering\cite{Kato22}.
Moreover, superconductivity has been observed in chiral crystals, 
Li$_{2}$Pd$_{3}$B and 
Li$_{2}$Pt$_{3}$B\cite{Bose05,Yuan06,Nishiyama07}, 
and its pairing mechanism due to electron-phonon coupling 
is elucidated\cite{Yuan06}.

In the theoretical aspect on the chiral phonons, 
the symmetry argument of the dielectric tensor\cite{Portigal68}, 
and the effective continuum field theory including 
rigid-body rotations\cite{Kishine20} 
based on the micropolar elastic theory\cite{Eringen12,Nowacki85} 
have been developed. 
The energy dispersions\cite{Boer96,Ghosh08}, 
acoustic activity in the $\alpha$ quartz\cite{Pine70} 
and the roton-like excitation in metamaterials\cite{Chen21} 
have been indeed observed.
On the other hand, the origin of characteristics of chiral phonons 
remains obscure, despite extensive studies based on 
microscopic models or first-principle 
calculations\cite{Pine70,Pine71,Teuchert74,Martin76,Chen22,Komiyama22}.
For example, what type of harmonic interactions cause
the energy splitting of chiral phonons with opposite CAM 
remains unknown.
In particular, as the splitting of acoustic phonons has 
a close relation to the stability of the given lattice structure, 
concrete constraint for chiral phonons 
is crucial for modeling real materials 
on the basis of the first-principle calculations.
Indeed, it is known that the first principles phonon calculations 
sometimes become unstable 
in chiral crystals\cite{Togo15,ABINIT,Phonon}.

In this letter, we have developed a microscopic theory 
on phonon energy dispersion in chiral crystals 
within a harmonic approximation, 
and elucidated some rigorous constraints 
and a source of splitting for opposite CAM modes.
One of the textbook examples of chiral crystals is Te\cite{Adenis89} 
and its phonon properties have also been studied 
both experimentally and 
theoretically\cite{Ghosh08,Pine70,Pine71,Teuchert74,Martin76,Chen22}.
So, let us first pick up this system and identify phonon properties 
characteristic to its chiral crystal structure.

\begin{figure}[b!]
\begin{center}
\includegraphics[width=8.5cm]{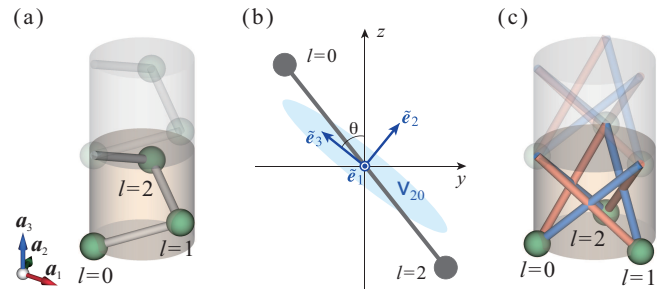}
\end{center}
\caption{
(Color online)
(a) A single helix in a Te-like lattice.
(b) Stiffness matrix $\sfv _{2,0}$ and its principle axes $\{ \tbee _j \}$.
(c) A double-handed triple helix.
Left and right bonds are colored in blue and red, respectively.
}
\label{fig1}
\end{figure}

At ambient pressure, Te has a crystal structure
categorized to the space group $P3_1 21$ (No.~152)
corresponding 
to the trigonal chiral point group $D_3$.
Its mirror image belongs to $P3_2 21$ (No.~154)
This structure is a hexagonal lattice made of helices 
running along the $z$-direction, 
and the unit cell contains three sublattice sites 
as shown in Fig.~\ref{fig1}(a).
While the primitive lattice vector 
$\baa _3 \Eq c (0,0,1)$ is along the $c$-axis, 
$\baa _1  \Eq a (1,0,0)$ and 
$\baa _2  \Eq a (-\frac12 , \frac{\sqrt3}{2},0)$ 
span the $ab$-plane, 
and we also define the supplementary vector 
$\baa _0 \Eq - \baa _1 \MINUS \baa _2$.
The site position of the three sublattices is then represented as 
\begin{equation}
\brr _l = \bss _l + \brr , \ \ 
\bss _l = \delta \, \baa _l + \FRAC13 l \, \baa _3 , \ \ 
(l = 0, 1, 2) ,
\label{eq:r_l}
\end{equation}
where $\brr \Eq \sum_{j=1}^3 n_j \baa _j$ with integer $n_j$'s 
is the unit cell position.
For Te, the value of $\delta$ is about 0.23\cite{Te-lattice}, 
but we let it be a free parameter $0 \LT \delta \LT \frac12$ 
and consider more general cases.
Site connectivity in each helix is described by 
the nearest-neighbor bonds 
$\btt _{l,l+1} \Equiv 
 \bss _{l+1} \MINUS \bss_l \PLUS \delta_{l,2} \msp2 \baa _3$ 
each connecting $l$ and $l \PLUS 1$ sublattices.
One should understand as 
$l \PLUS 1 \Eq 3 \Equiv 0$ as well as 
$l \MINUS 1 \Eq -1 \Equiv 2$ 
for the sublattice index throughout this paper.

Now we summarize the procedure of calculating phonon energy dispersion.  
A starting point is a lattice deformation energy functional 
$\mathscr{V}$ represented in terms of atomic displacements 
$\{ \bdd (\brr _l ) \}$, 
and it is customary for it to employ a quadratic form of 
$\{ \bdd (\brr _l ) \}$.
After Fourier transformation, 
the coefficient in the quadratic form for each wavevector $\bkk$ 
is reduced to a matrix with the dimension of 
$3 \msm1 \times \msm1 \mbox{(no.~of sublattices)}$.
This is called the dynamical matrix, 
and its each eigenvalue $\lambda_{\alpha \bkk}$ 
determines the corresponding phonon energy dispersion as 
$\hbar \omega_{\alpha \bkk} \Eq 
 \hbar \sqrt{\lambda_{\alpha \bkk}/M}$ 
with the atomic mass $M$.
Since our concern is their $\bkk$-dependence, 
we use the units of $M \Eq 1$ and $\hbar \Eq 1$.
In this paper, we treat for simplicity the cases 
that constituent atoms are all identical 
and they have an isotropic mass tensor, 
but it is straightforward to generalize these points.

Let us then construct a simple model for the deformation energy 
$\mathscr{V}$ describing two-atom interactions as explained before.
To this end, two requirements are crucial.
First, its energy must be
non-negative for any displacements 
to guarantee the stability of the equilibrium structure.
Secondly, its functional form must match the lattice symmetry.
Namely, it should be invariant upon any symmetry operation of the lattice.
Considering $\mathscr{V}$ is scalar, 
a product of two displacement vectors 
$\bdd \Eq \bdd (\brr _l )$ and 
$\bdd ' \Eq \bdd (\brr _m )$ 
appears in it with a coefficient that is either a scalar ($v$), 
symmetric second-rank tensor ($v_{i \msm1 j} \Eq v_{j \msp1 i}$), 
or antisymmetric tensor ($v_{ij} \Eq \epsilon_{i \msm1 j \msp1 k} D_k$).
The last case corresponds to an interaction of 
Dzyloshinskii-Moriya (DM) type $\bDD \cdot (\bdd \times \bdd ')$, 
but this type is not allowed in our case.
Clearly, it does not fulfill the first requirement, 
and one cannot resolve this problem by generalizing its form.
The remaining two cases are all together summarized to a quadratic form 
$(\bdd \MINUS \bdd ') \cdot \sfv  \, (\bdd \MINUS \bdd ')$ 
with a real symmetric {\DIM3} matrix 
$\sfv  \Eq \sfv (\brr _{l} \MINUS \brr  _m)$, 
which will be referred to as \textit{stiffness matrix} henceforth.
This complies with the first requirement, 
if $\sfv $ has no negative eigenvalue.
Furthermore, energy cost is zero when $\bdd \Eq \bdd '$, 
which meets our expectation for $\mathscr{V}$ 
since the related atomic bond is intact.
The second requirement imposes that $\sfv $'s principal axes 
$\{ \tbee _i \}$ ($i \Eq 1, 2, 3$) 
point to the local symmetric directions of the bond 
connecting the atomic sites $\brr _l$ and $\brr _m$.

We are ready to write down an explicit form of $\mathscr{V}$ 
for the case of Te-like lattice with ``left'' handedness.
It consists of a part for inside each helix ($\mathscr{V}_{L,1}$) 
and a part between neighboring helices ($\mathscr{V}_{L,2}$). 
The latter part will be discussed later.
For $\mathscr{V}_{L,1}$, 
we consider contributions of only nearest-neighbor pairs
\begin{equation}
\mathscr{V}_{L,1} = \! \sum_{\brr _l} 
\bigl[ \bdd (\brr _l ) \MINUS 
       \bdd (\brr _l \PLUS \btt _{l,l+1} ) \bigr] \cdot 
\sfv _{l,l+1} \, 
\bigl[ \bdd (\brr _l ) \MINUS 
       \bdd (\brr _l \PLUS \btt _{l,l+1} )\bigr] , 
\label{eq:VL1}
\end{equation}
where $\sfv _{l,l+1}$ is the stiffness matrix 
for the bond $\btt _{l,l+1} $ 
and is subject to the symmetry constraints 
$\sfv _{2,0} \Eq 
 \sfC _3   \sfv _{0,1} \sfC _3^T \Eq 
 \sfC _3^T \sfv _{1,2} \sfC _3  $.
Here, $\mathsf{C}_3$ is the {\DIM3} matrix of clockwise rotation 
by the angle $\FRAC{2\pi}{3}$ about the $z$-axis, 
and the symbol $T$ denotes a matrix transposition.
As for the bond $\btt _{2,0}$, 
it remains intact under the $\pi$-rotation about the $x$-axis $C_{2x}'$, 
and therefore one principle axis needs to be 
$\tbee _1 \Eq (1,0,0)^T$.
This leads to a parametrization for the other axes as 
$\tbee _3 \Eq (0, -s_\theta , c_\theta )^T$ and 
$\tbee _2 \Eq (0,  c_\theta , s_\theta )^T$ 
with short hand notations 
$s_\theta \Eq \sin \theta$ and $c_\theta \Eq \cos \theta$, 
but the lattice symmetry provides no further constraints 
to the $\theta$ value (see Fig.~\ref{fig1}(b)).
Nonetheless, it is reasonable to expect that 
one local axis is close to the bond direction 
$\tbee _3 \msm2 \sim \msm2 \btt _{2,0}/|\btt _{2,0}|$.
If they coincide, the value is then 
$\theta \Eq \tan ^{-1} (\! \sqrt{27} \msp2 \delta \msp2 a/c)$.
For the $\btt _{2,0} $ bond, 
using its local axes $\tbee _i$'s discussed above, 
we can write down its stiffness matrix as
\vspace{-2pt}
\begin{equation}
\sfv _{2,0} = \sum_{i=1}^3 K_i^{} \, \tbee _i^{} \, \tbee _i^T 
= 
\left[
\begin{array}{@{}c@{\hs{1mm}}c@{\hs{1mm}}c@{}}
K_{11} & 0 & 0 \\ 
0 & \phantom{-} K_{22} & -\varDelta K \\
0 & -\varDelta K & \phantom{-} K_{33} 
\end{array}
\right] 
=: \bsfv , 
\label{eq:v20}
\vspace{-4pt}
\end{equation}
where $K_i$'s are non-negative stiffness constants, 
and $K_{11}^{} \Eq K_1^{}$,
$K_{22}^{} \Eq K_2^{} c_{\theta}^2 \PLUS K_3^{} s_{\theta}^2$, 
$K_{33}^{} \Eq K_2^{} s_{\theta}^2 \PLUS K_3^{} c_{\theta}^2$, and
$\varDelta K \Eq 
(K_3^{} \MINUS K_2^{} ) s_{\theta}^{} \msp2 c_{\theta}^{}$.
The other stiffness matrices $\sfv _{0,1}$ and $\sfv _{1,2}$ 
can be calculated by multiplying $\sfC _3^{}$ and $\sfC _3^T$ 
to this $\sfv _{2,0}^{}$.

The next step is Fourier transformation and 
we obtain the dynamical matrix.
We define a 9-dimensional super-vector 
by combining the displacement vectors in the wavevector $\bkk$-space 
for the three sublattices 
$\vec{d}(\bkk ) \Equiv 
 (\bdd _0 (\bkk ), \, \bdd _1 (\bkk ), \, \bdd _2 (\bkk ) )^T$.
The intra-helix energy then reads as 
$\mathscr{V}_{L,1} \Eq 
 \sum_{\bkk} \vec{d}(-\bkk ) \cdot \ulV _{L,1} (\bkk ) 
 \vec{d} (\bkk )$, 
and its coefficient $\ulV _{L,1} (\bkk )$ is the dynamical matrix.
It is a {\DIM9} hermitian matrix and
\begin{equation} 
\hs{-0.6mm}
\ulV _{L,1} (\bkk ) 
\msm2 = \msm2
\left[
\begin{array}{@{}c@{\hs{1mm}}c@{\hs{1mm}}c@{}}
  \sfC _3^T \msp1 \bsfv \msp2 \sfC _3 \PLUS 
  \bsfv & 
- \sfC _3^T \msp1 \bsfv \msp2 \sfC _3 \gamma_{0,1} & 
- \bsfv \gamma_{2,0}^*
\\[4pt]
- \sfC _3^T \msp1 \bsfv \msp2 \sfC _3 \gamma_{0,1}^* & 
  \sfC _3   \msp1 \bsfv \msp2 \sfC _3^T \PLUS
  \sfC _3^T \msp1 \bsfv \msp2 \sfC _3 & 
- \sfC _3   \msp1 \bsfv \msp2 \sfC _3^T \gamma_{1,2}
\\[4pt]
- \bsfv \gamma_{2,0} & 
- \sfC _3   \msp2 \bsfv \msp2 \sfC _3^T \gamma_{1,2}^* &
  \bsfv \PLUS \sfC _3 \msp2 \bsfv \msp2 \sfC _3^T 
\end{array}
\right], 
\label{eq:VL1k}
\end{equation}
where the phase factors 
$\gamma_{l,l+1} \Equiv 
 \exp (i \bkk \msm2 \cdot \msm2 \btt _{l,l+1})$ 
depend on $\bkk$.

In the following we consider the case of $\bkk \Eq (0,0,k_z)$ 
parallel to the screw axis and denote the wavevector dependence 
by $k_z$ alone.
In this case, 
concerning the interaction energy 
between neighboring helices $\sV _{L,2}$, 
its dominant terms simply renormalize the parameters in $\bsfv$ 
in Eq.~\eqref{eq:VL1k}.  
Therefore, it suffice to analyze $\ulV _{L,1} (k_z )$ 
with $\bsfv$ which should be understood as a renormalized coupling.  
Details will be published elsewhere.
Then, the three phase factors coincide as 
$\gamma_{l,l+1} \Eq e^{ip}$ with 
$p \Eq \FRAC13 k_z c$, 
and $\ulV _{L,1} (k_z )$ has a special symmetry.
There exists an orthogonal matrix 
$\ulCCC$ that commutes with 
$\ulV _{L,1} (k_z )$.
It is defined by the transformation of the sublattice displacements as 
$\ulCCC : {\bdd _{l}} (k_z ) \Eq 
 \sfC _3 \bdd _{l+1} (k_z )$ for all $l$'s.
Since this satisfies the relation $\ulCCC ^3 \Eq \ulone$, 
its eigenvalues are $\zeta^{m}$ ($m =0, \pm 1$) 
with $\zeta \Eq e^{\msp2 i2\pi/3}$, 
and the eigenspaces of $\ulV _{L,1} (k_z)$ are split into 
the three subspaces each with a different $m$ value.
This $m$ is precisely the quantum number of CAM 
mentioned in the introduction. 
For later use, we introduce a hermitian operator that determines 
the value of CAM: 
$\underline{L}_z \equiv 
(\underline{C}_3 \MINUS \underline{C}_3^{\msp2 T} )/(\sqrt3 \msp2 i) $.
It is easy to check that its eigenvalues are $m=0$ and $\pm 1$. 

Our task is thus reduced to diagonalizing 
an effective dynamical matrix in each CAM subspace.
It is a {\DIM3} hermitian matrix given as
\vspace{-6pt}
\begin{equation} 
\sfv _{\mathrm{eff}}^{(m)}  (k_z )
= 
\sfF \bigl( \phi_{m} ( k_z ) \bigr)
 \msp2 \bsfv \msp4 
\sfF \bigl( \phi_{m} ( k_z ) \bigr) , 
\label{eq:veff}
\end{equation}
with $\phi_{m} (k_z) \Equiv \FRAC13 ( 2 m \pi \PLUS k_z c )$ 
and its eigenvector is the sublattice component $\bdd _1 (k_z )$.
The other components can be obtained using the relation, 
$\bdd _{l\pm 1}^{} (k_z ) \Eq 
 \zeta_{}^{ \pm m} \msp2 
 \sfC _3^{\mp 1} 
 \bdd _l^{} \msp2 (k_z )$.
Here, 
$\sfF (P) \Equiv i \msp2 
 ( \msp2 e^{iP/2} \msp2 \sfC _3 \MINUS 
   e^{-iP/2}\msp2 \sfC _3^T \msp2 )$ 
is hermitian, and
$\sfF (-P) \Eq - \sfF (P)^*$.
The expression \eqref{eq:veff} immediately demonstrates 
an important chiral symmetry 
$\sfv _{\mathrm{eff}}^{(-m)}  (-k_z ) \Eq 
 \sfv _{\mathrm{eff}}^{(m)}  (k_z ) ^* $, 
which is a consequence of the time reversal symmetry of 
the energy $\mathscr{V}$. 
This guarantees a common set of eigenvalues for 
$\sfv _{\mathrm{eff}}^{(+1)}  (k_z ) $ and 
$\sfv _{\mathrm{eff}}^{(-1)}  (-k_z )$, 
which may be called a chiral pair.

\begin{figure}[t!]
\begin{center}
\includegraphics[scale=0.17]{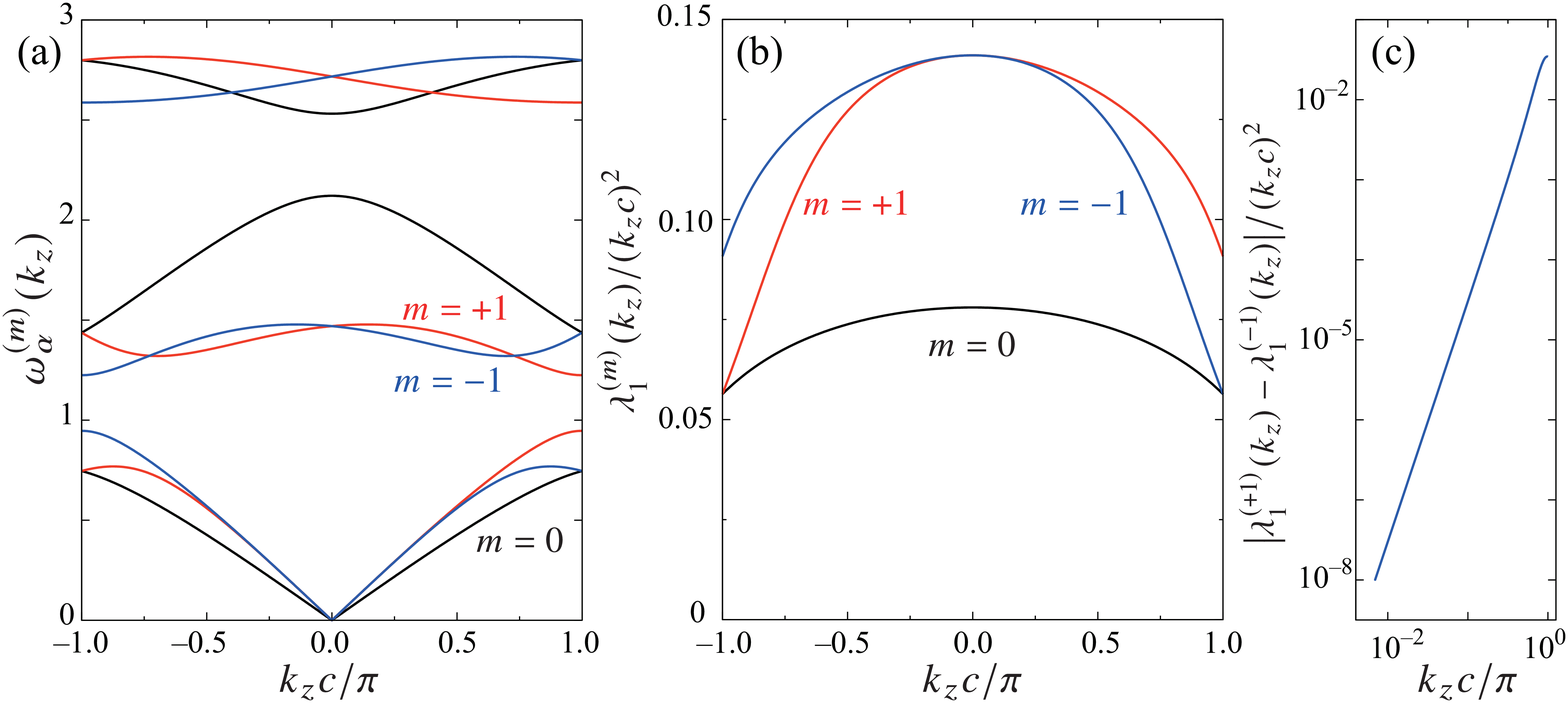}
\end{center}
\vspace{-10pt}
\caption{
(Color online)
(a) Phonon energy dispersion in a Te-like chiral lattice. 
Different CAM subspaces are distinguished by color.
(b) $\lambda_{1}^{(m)}(k_z )/(k_z c)^2$.  
(c) Scaling of the splitting 
$\bigl| \lambda_{1}^{(+1)}(k_z )-\lambda_{1}^{(-1)}(k_z ) \bigr|$. 
The slope indicates this scale as $k_z^5$.  
}
\vspace{-12pt}
\label{fig2}
\end{figure}

We have numerically diagonalized 
$\sfv _{\mathrm{eff}}^{(m)}(k_z )$'s 
for each $k_z$ to obtain their eigenvalues 
$\lambda_{\alpha}^{(m)} (k_z)$ 
($\alpha \Eq 1,2,3$ in ascending order), 
and determined phonon energies 
$\omega_{\alpha}^{(m)} (k_z ) \Eq 
 [ \lambda_{\alpha}^{(m)} (k_z)]^{1/2}$.
Their energy dispersions are plotted in Fig.~\ref{fig2}(a) 
for the set of parameters $K_3 \Eq 2 K_1 \Eq 6 K_2 = 3.0$ 
and $\theta \Eq 0.3\pi$.
Each CAM subspace has one acoustic ($\alpha \Eq 1$) 
and two optical ($\alpha \Eq 2, 3$) modes.
The degeneracy is a little complicated at the Brillouin zone (BZ) boundary 
$k_z \Eq \pm \pi/c \equiv \pm k_0$, since the lattice 
is nonsymmorphic.~\cite{Hernandez20,Hernandez21}  
As the effective ``phase'' $\phi_m (k_z)$ in Eq.~\eqref{eq:veff}
is connected through different CAM subspaces, 
this results in 
$\omega_{\alpha}^{(-1)} (k_0) \Eq 
 \omega_{\alpha}^{(0)} (-k_0) \Eq 
 \omega_{\alpha}^{(0)} (k_0) \Eq 
 \omega_{\alpha}^{(+1)} (-k_0)$ 
and 
$\omega_{\alpha}^{(+1)} (k_0) \Eq 
 \omega_{\alpha}^{(-1)} (-k_0)$.   
Comparing the $m \Eq \pm 1$ subspaces, 
the energy of the acoustic mode 
is nearly degenerate around $k_z \Eq 0$.

\vspace{-10pt}

There have been some discussions about a possibility 
of sound velocity splitting between the $m \Eq \pm 1$ CAM 
subspaces\cite{Portigal68}.
However, if the velocity value were split in the $k_{z} \Eq 0$ limit, 
it would require a nonanalyticity of 
$\lambda_{1}^{(\pm 1)} (k_z)$, 
which one can easily disprove.
In each $m$ CAM subspace, 
$\sfv _{\mathrm{eff}}^{(m)}  (k_z )$ 
is a smooth function of $k_z$, 
and its three eigenvalues are well separated at $k_z \Eq 0$.
These two mean that the perturbation in $k_z$ should have 
a nonvanishing convergence radius,
and disprove the nonanalyticity.  
Numerical analysis of our data concludes 
$\lambda_{1}^{(\pm 1)} (k_z) \msm2 \sim \msm2
 u_{}^2 \msp1 k_z^2 \pm  
 w_5^{} \msp1 k_z^5 $ as shown in Fig.~\ref{fig2}(b) and (c). 
As for phonon energy, this leads to the asymptotics
$\omega_{1}^{(\pm 1)} (k_z) \msm2 \sim \msm2
 |k_z | \msp2 
 \bigl[ u \pm ( w_5 / 2u ) \msp2 k_z^3  \bigr ]$ 
with the sound velocity $u$, 
and thus an energy splitting appears 
in the order $O(k_z^4)$. 
Note that in more general models, 
corrections may start from a lower-order such as $O(k_z^3)$, 
and then it leads to an energy splitting of $O(k_z ^2)$.  

Energy splitting due to chiral structure is more evident 
for the optical modes ($\alpha \Eq 2,3$).
It appears in the order $O(k_z)$ as
\begin{equation}
\lambda_{\alpha}^{(m)}(k_z) \sim 
 \lambda_{\alpha} (0) \PLUS 
 \varGamma_{\alpha} \msp2 k_z \msp2 m . 
\end{equation}
Its coefficient $\varGamma_{\alpha}$
is an important indicator that quantifies 
the effects of chiral structure on phonon dispersion, 
and we will call it \textit{splitting coefficient}.
Since $k_z$ is a polar vector element and $m$ is axial, 
the splitting coefficient $\varGamma_a$ needs to be a pseudo-scalar. 
It should change sign under $z$-plane mirror operation 
but remains invariant under time reversal.  
In order to see which parts of the chiral structure dominate 
$\varGamma _\alpha$'s, 
it is desirable to realize a smooth control of chiral lattice 
structure and how it affects the values of $\varGamma _\alpha$'s.

Such a control of lattice chirality and its handedness 
is actually accomplished 
by glueing a chiral crystal together with its mirror image, 
i.e., \textit{enantiomorph}.
In the present case, one realization is the lattice 
with the unit cell shown in Fig.~\ref{fig1}(c), 
which we will call \textit{double-handed triple helix} (DHTH).
This contains three sublattice sites positioned 
at $\brr _l$ in Eq.~\eqref{eq:r_l} 
but now with $\bss _l \Eq \delta \baa _l$, 
and then two types of bonds are assigned for opposite windings
$\btt _{l,l \pm 1} \Eq 
 \bss _{l \pm 1} \MINUS \bss _{l}
  \PLUS \baa _3$.
While the deformation energy of 
the \textit{left-hand} wound ($L$) bonds is 
given by Eq.~\eqref{eq:VL1} with this redefinition, 
the right-hand wound ($R$) bonds contribute its counterpart 
$\mathscr{V}_{R,1} \Eq \! 
 \sum_{\brr _{l}} 
 \bigl[ 
    \bdd (\brr _l ) \MINUS \bdd (\brr _l \PLUS \btt _{l,l-1} ) 
 \bigr] \cdot 
 \sfv _{l,l-1} 
 \bigl[ 
    \bdd (\brr _l ) \MINUS \bdd (\brr _l \PLUS \btt _{l,l-1} )
 \bigr]$.
To realize a completely nonchiral limit, one sets as 
$\sfv _{l+1,l} \Eq 
 \sfm _z \msp2 \sfv _{l,l+1} \msp2 \sfm _z$ 
with the mirror operation 
$\sfm _z \Equiv \mbox{diag }(1,1,-1)$, 
since $\btt _{l+1,l} \Eq - \sfm _z \msp2 \btt _{l,l+1}$.
In particular, $\hsfv \Equiv \sfv _{0,2}$ is identical to 
$\sfv _{2,0}$ in Eq.~\eqref{eq:v20} 
except for the plus sign for $\varDelta K$.
However, in more general cases 
in which handedness cancellation is incomplete, 
the parameters $K_{jj}$ and $\varDelta K$ differ 
between $L$- and $R$-bonds, 
and henceforth, we will add the label $L$ or $R$ to distinguish them.
As before, the symmetry constraint determines 
the stiffness matrix for the other bonds 
$\sfv _{0,2}^{} \Eq 
 \sfC _3^{} \msp2 \sfv _{1,0}^{} \msp2 \sfC _3^T \Eq 
 \sfC _3^T  \msp2 \sfv _{2,1}^{} \msp2 \sfC _3^{}$.

\begin{figure}[t!]
\begin{center}
\includegraphics[scale=0.17]{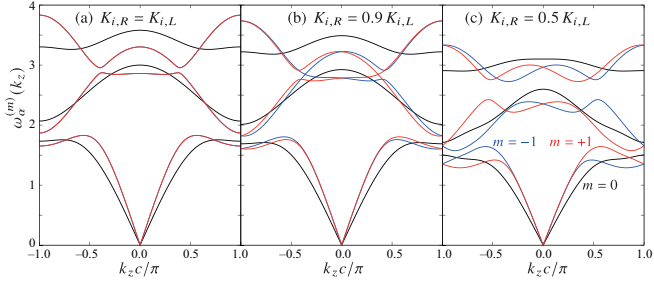}
\end{center}
\vspace{-10pt}
\caption{
(Color online)
Phonon energy dispersion in the DHTH lattice. 
The ratio $K_{i,R}/K_{i,L}$ is common for all $i$'s.  
The $m \Eq \pm 1$ CAM modes are degenerate in (a). 
}
\vspace{-12pt}
\label{fig3}
\end{figure}

In a general DHTH lattice, the total deformation energy is 
$\mathscr{V}_{L,1} \PLUS \mathscr{V}_{R,1}$, 
and this leads to the dynamical matrix 
$\ulV _{L,1} (\bkk )\PLUS \ulV _{R,1} (\bkk ) 
 \msp2 \equiv \ulV (\bkk ) $.
As before, we will concentrate on the case of $\bkk \Eq (0,0,k_z)$.
Then, $\ulV _{R,1} (k_z ) \Eq \ulV _{R,1} (0,0,k_z )$ 
is given by modifying the form in Eq.~\eqref{eq:VL1k}: 
first replace $\bsfv$ by $\hsfv$, 
and then interchange $\sfC _3$ and $\sfC _3^T$, 
and lastly operate complex conjugation.
Since this also commutes with $\ulCCC$ defined before, 
$\ulV _{R,1} (k_z)$ is reduced to a {\DIM3} effective matrix 
$\sfv _{\mathrm{eff},R}^{(m)}(k_z) \Eq 
 \sfF ( \phi_{m}^{-} ) \msp2 \hsfv \msp4 
 \sfF ( \phi_{m}^{-} )$, 
while its counterpart is 
$\sfv _{\mathrm{eff},L}^{(m)}(k_z) \Eq 
 \sfF ( \phi_{m}^{+} ) \msp2 \bsfv \msp4 
 \sfF ( \phi_{m}^{+} )$.
Here, 
$\phi_{m}^{\pm} \Equiv k_z c \pm \FRAC{2m\pi}{3}$ 
and it has a different factor of $k_z c$
from the one in Eq.~\eqref{eq:VL1} due to the redefinition of 
the bond vectors $\btt _{l,l\pm 1}$.
The total effective dynamical matrix is 
$\bsfv _{\mathrm{eff}}^{(m)}(k_z) \Equiv 
 \sfv _{\mathrm{eff},L}^{(m)}(k_z) \PLUS 
 \sfv _{\mathrm{eff},R}^{(m)} (k_z)$, 
and it is instructive to rewrite it into the following form
\begin{equation} 
\bsfv _{\mathrm{eff}}^{(m)} \msp1 (k_z )  = 
\bsfv _0^{} - 
\zeta_{}^{\msp2 -m} \msp2 
\sfC _3^T \msp2 
\sfv _s^{} (-p) \msp2 
\sfC _3^T - 
\zeta_{}^{\msp2 m} \msp2 
\sfC _3^{} \msp2  
\sfv _s^{} ( p) \msp2 
\sfC _3^{} , 
\end{equation}    
where 
$p \Eq k_z c$.
The constant term is 
$\bsfv _0 \Equiv 
 \FRAC12 \msp2 \mbox{diag } 
 \bigl( 
      K_{11,+} \msm4 \PLUS \msm2 3 K_{22,+}, 
    3 K_{11,+} \msm4 \PLUS \msm2   K_{22,+}, 
    4 K_{33,+} \msm2 
 \bigr) \PLUS 
 \varDelta K_{-} \sfX$ 
defined with 
$(\sfX )_{\mu,\nu} \Equiv 
 \delta_{\mu,y} \delta_{\nu,z} \PLUS 
 \delta_{\mu,z} \delta_{\nu,y}$ 
and the (anti-)symmetrized parameters 
$K_{jj,\pm} \Equiv K_{jj,L} \pm K_{jj,R}$ and 
$\varDelta K_{\pm} \Equiv \varDelta K_{L} \pm \varDelta K_{R}$.
The $k_z$-dependence comes from 
$\sfv _s (p) \Eq 
 \cos p \msp2 \sfv _+ \PLUS i \sin p \msp2 \sfv _-$ 
defined with 
$\sfv _{\pm } \Eq \mbox{diag } 
 \bigl( K_{11,\pm}, K_{22,\pm}, K_{33,\pm} \bigr) \MINUS 
 \varDelta K_{\mp} \sfX$.
Note that $\sfv _s (p)$ is not hermitian, 
but has the relation 
$\sfv _s (-p) \Eq \sfv _s (p)^*$. 
As in the previous case of single helix, 
the time reversal symmetry leads to the relation 
$\bsfv _{\mathrm{eff}}^{(-m)}  (-k_z ) \Eq 
 \bsfv _{\mathrm{eff}}^{(m)}  (k_z ) ^{*}$, 
and this guarantees the energy degeneracy of chiral phonons 
$\omega _{\alpha}^{(-m)} (-k_z ) \Eq 
 \omega _{\alpha}^{(m)} (k_z )$, 
in general DHTH lattices as shown in Fig.~\ref{fig3}. 
The degeneracy at the BZ boundary 
$k_z \Eq \pm k_0$ 
differs from the behavior 
in Fig.~\ref{fig2}(a). 
As the DHTH lattice is now symmorphic, 
there holds the relation 
$\bsfv _{\mathrm{eff}}^{(m)}  (k_0 ) \Eq 
 \bsfv _{\mathrm{eff}}^{(m)}  (-k_0 ) \Eq 
 \bsfv _{\mathrm{eff}}^{(-m)}  (\pm k_0 ) ^{*}$, 
and this leads to the degeneracy 
$\omega _{\alpha}^{(+1)} (k_0 ) \Eq 
 \omega _{\alpha}^{(+1)} (-k_0 ) \Eq 
 \omega _{\alpha}^{(-1)} (\pm k_0 )$. 
Note that when $\mathscr{V} _{R,1}$ is set to zero, 
the system is reduced to decoupled three copies of a Te-like lattice 
but their lattice constant $c$ is tripled.

\vspace{-10pt}

It is important to examine the nonchiral limit 
where the right-handed part is a precise mirror image of the left-handed part.
In this case, the antisymmetrized parameters vanish 
($K_{jj,-} \Eq \varDelta K_{-} \Eq 0$), 
and this leads to the special symmetries that 
$\sfv _s (-p) \Eq \sfm _z \msp2 \sfv _s (p) \msp2 \sfm _z$ 
and $\bsfv _0$ is a diagonal matrix.
Combining them with the property 
$ [ \sfC _3 ^{\pm1}, \msp2 \sfm _z ] \Eq 0$, 
it is easy to show 
$\bsfv _{\mathrm{eff}}^{(m)}  (-k_z ) \Eq 
 \sfm _z \msp2 \bsfv _{\mathrm{eff}}^{(m)} (k_z ) \msp2 \sfm _z$.
Together with the chiral symmetry, 
this leads to the degeneracy of phonon energy dispersion 
between the opposite CAM modes, 
$\omega _{\alpha}^{(m)} ( k_z ) \Eq \omega _{\alpha}^{(-m)} ( k_z )$.
Note that this additional symmetry is specific to the nonchiral limit.

Finally, let us evaluate the energy splitting 
for the opposite CAM modes, 
which determines the splitting coefficient
$\varGamma _\alpha$ introduced before.
We can calculate this by following the standard procedure 
of the first-order perturbation in $k_z$.
Let $\vec{d}_\alpha^{\msp4 (m)}$ denote the 9-dimensional eigenvectors 
of $\underline{V} (k_z \!=\! 0)$ 
in the CAM subspaces $m \!=\! \pm 1$. 
They satisfy the relation 
$\vec{d}_\alpha^{\msp4 (-1)} \Eq 
[ \vec{d}_\alpha^{\msp4 (+1)} ]^{\, *}$. 
We can show immediately 
$m \msp2 \varGamma_\alpha \Eq 
\langle \vec{d}_\alpha^{\msp4 (m)} \msp2 | \msp2 
\underline{D}_z \msp2 | \msp2
\vec{d}_\alpha^{\msp4 (m)} \rangle$ 
with 
$ \underline{D}_z \equiv 
\bigl[ d \underline{V} (k_z)/d k_z \bigr]_{k_z =0}$, 
and this new matrix is represented as 
\begin{equation}
\underline{D}_z
= -i \left[
\begin{array}{@{\hs{2pt}}c@{\hs{4pt}}c@{\hs{4pt}}c@{\hs{2pt}}}
\mathsf{0} &
z_{01} \msp2
\mathsf{C}_3^T \msp2 \mathsf{v}_{-} \msp2 \mathsf{C}_3  &
z_{02} \msp2         \mathsf{v}_{-} 
\\[2pt]
z_{10} \msp2
\mathsf{C}_3^T \msp2 \mathsf{v}_{-} \msp2 \mathsf{C}_3 &
\mathsf{0} &
z_{12} \msp2
\mathsf{C}_3   \msp2 \mathsf{v}_{-} \msp2 \mathsf{C}_3^T 
\\[2pt]
z_{20} \msp2         \mathsf{v}_{-} &
z_{21} \msp2
\mathsf{C}_3 \msp2   \mathsf{v}_{-} \msp2 \mathsf{C}_3^T &
\mathsf{0} 
\end{array}
\right] . 
\label{eq:matrixJz}
\end{equation}
Here, $z_{l l'}$ is the $z$-component of 
the bond vector $\btt _{l l'}$, and 
$z_{l' l} \Eq -z_{l l'}$.  
Thus, this hermitian matrix $\underline{D}_z$ is pure imaginary 
and similar 
to a current in the sense that it changes sign under 
each of time reversal and space inversion operations.  
It is possible to separate the pseudo-scalar part $\varGamma_a$ alone 
by combining the CAM operator: 
\begin{align}
\varGamma_\alpha \Eq 
 \langle \vec{d}_\alpha^{\msp4 (m)} \msp2 |  \msp2 
 \underline{\varGamma} \msp2 | \msp2 
 \vec{d}_\alpha^{\msp4 (m)} \rangle , \ \ 
\mbox{with } 
\underline{\varGamma} \equiv 
\underline{L}_z \, \underline{D}_z.
\label{eq:gammaa}
\end{align}
Note that $\underline{\varGamma}$ is real symmetric, 
since $\underline{L}_z$ and $\underline{D}_z$ commute. 
Thus, this $\underline{\varGamma}$ is the operator 
that defines the chirality (handedness) of the system.  
We skip the detail of calculating the matrix element $\varGamma_\alpha$ 
and show the final result: 
\begin{equation}
\varGamma _{\alpha} = 
\FRAC{\sqrt3}{2} \, 
\Bigl[
        \bigl(1 \mp \cos \beta \bigr) K_{33,-} \MINUS 
\FRAC12 \bigl(1 \pm \cos \beta \bigr) \bigl( K_{11,-} \PLUS K_{22,-} \bigr)
\Bigr], 
\end{equation}
where the upper and lower signs are for 
$\alpha \Eq 2$ and 3, respectively, and 
$\beta \Equiv  \tan^{-1} 
 \bigl[ 
   \sqrt8 \msp2 \varDelta K_{-} 
   ( K_{11,+} \msm2 \PLUS K_{22,+} \msm2 \MINUS 2 K_{33,+} )^{-1}\bigr]$.
It is interesting that they satisfy the simple sum rule 
$\varGamma_2 \PLUS \varGamma_3 \Eq  \sqrt3 \msp2 
 \bigl[ K_{33,-} \MINUS 
       \FRAC12 \bigl( K_{11,-} \PLUS K_{22,-} \bigr) 
 \bigr] \Equiv \varGamma_s $, 
which is also represented in terms of the tensor's principle values as 
$\varGamma _s \Eq 
 \FRAC14 (K_{2,-} \PLUS K_{3,-} \MINUS 2 K_{1,-}) \PLUS 
 \FRAC34 (K_{3,-} \MINUS K_{2,-}) \cos 2 \theta $, 
where $\theta$ is the effective bond angle introduced before
for defining $\sfv _{2,0}$.
This $\varGamma_{s}$ is completely determined 
by the antisymmetrized parameters 
(i.e., imbalance between the left- and right-handed parts), 
and independent of the symmetrized parameters 
(i.e., common factors in the two parts).
Therefore, we may consider $\varGamma _s$ 
as the most fundamental indicator characterizing 
the effects of chiral structure on phonon dispersion.
The parameter $\beta$ determines how the sum 
$\varGamma _s$ is distributed between the two optical modes.
Note that one can apply these results 
to the systems with one chiral part alone 
such as Te lattice, 
and in those cases 
$ K_{jj,+} \Eq |K_{jj,-} |$ and 
$ |\varDelta K_{+}| \Eq |\varDelta K_{-}|$.
It is clear that $\varGamma_s$ changes its sign 
upon a reversal of lattice handedness.

It is important to notice that one can represent 
this indicator as an averaged uniaxial anisotropy 
$Q_u (\sfv ) \Eq \! \FRAC{\sqrt3}{4} \, 
 \mbox{Tr} \msp2 [ (1 \MINUS 3 \sfm _z ) \, \sfv ]$
of the local stiffness matrices 
weighted by the bond sign of handedness 
$ h \Eq +1$ for left and $-1$ for right bonds: 
$\varGamma _s \Eq 
 N^{-1} \sum_{\bRR}
 h (\bRR ) \, 
 Q_u (\sfv (\bRR )) $ with 
$N$ being the number of sites. 
$\bRR$ denotes a bond center position measured from the nearest 
chiral axis, 
and if multiple $\sfv$'s exist at the same $\bRR$, 
they should be all counted.  

Since the splitting coefficients $\varGamma_{\alpha}$ are 
a kind of the ``order parameters'' 
of the chiral system, they belong to a nontrivial representation 
of the supergroup $D_{3h}$. 
Considering $\underline{L}_{z}$ and $\underline{D}_{z}$ 
in Eq.~\eqref{eq:gammaa} belong to $A_{2}'$ and $A_{2}\dprime$, respectively, 
$\varGamma_{\alpha}$'s have the same symmetry 
as $m \cdot k_z$, i.e.,
$A_2 ' \otimes A_2\dprime \Eq A_1\dprime$.  
In the language of cluster multipoles, 
they correspond to the pseudo-scalar (electric toroidal) 
multipoles of $G_{0}$ and $G_{u}$-type~\cite{Kishine22,Hayami18}, 
the latter of which gives the mono-axial anisotropy. 
A caution is necessary for the term ``pseudo-scalar'' 
in the systems without inversion or any mirror symmetries.
It is instructive to elevate the system's symmetry 
by supplementing generator(s) of mirror or inversion type.  
In our case, 
with the mirror $\sfm _z$ supplemented, 
the chiral point group $D_3$ is
elevated to its nonchiral supergroup $D_{3h}$,  
which is the symmetry of the DHTH with the symmetric couplings 
$K_{j,R} \Eq K_{j,L}$.
Pseudo-scalars are defined as bases of its $A_1\dprime$ representation, 
which changes sign under improper rotations and mirror operations.
When the system is chiral ($K_{j,R} \neq K_{j,L}$), 
$\underline{\varGamma}$ falls into the identity representation $A_{1}$ 
of the $D_{3}$ point group, 
and gives a nonvanishing contribution to the CAM mode splitting.
Similarly, in the cubic chiral systems, 
the lowest-order pseudo-scalar multipoles
($A_{1u}$, $A_{2}$, $A_{u}$ in the supergroup $O_{h}$, $T_{d}$, $T_{h}$) 
of $G_{0}$ and $G_{4}$-type\cite{Kishine22,Hayami18} 
induce a CAM splitting.

To summarize, we have developed a microscopic theory 
on the energy dispersion of chiral phonons
within the harmonic approximation.
Those chiral phonons are characterized by the crystal angular momentum 
$m \Eq \pm 1$, and the splitting of 
their energy dispersions depending on $m$. 
This $k_z$-linear energy splitting in the optical modes 
is indicated by nonvanishing splitting coefficients $\varGamma_{\alpha}$. 
They are thus order parameters of the chiral system, 
belonging to the nontrivial $A_{1}\dprime$ representation 
of the supergroup, and related to 
the $G_0$- and $G_u$-type electric toroidal multipoles 
(or $G_4$ type when the chiral system is cubic).  
Analyticity of the dynamical matrix in $\bkk$ enforces
an identical value of sound velocity for their acoustic modes,  
and a splitting appears in the order of at least $k^{2}$ or higher.
It is also important that stability of the equilibrium structure
forbids the presence of antisymmetric interactions, 
because they otherwise violate
the positivity of stiffness matrices.
These constraints are crucial for modeling real materials 
in the first-principles phonon calculations. 
The splitting is much more visible in optical modes, 
and it starts from the order $O(k^1)$.  
Its size is determined by the uniaxially anisotropic component 
of the stiffness matrices about the chiral axis. 
These fundamental findings of the present paper will provide
further insights for elucidating the phonon and 
related phenomena in chiral systems.

\vspace{3mm}
\begin{acknowledgment}

The authors thank Jun-ichiro Kishine, Hiroyasu Matsuura, and Kazumasa Hattori 
for fruitful discussions.
This research was supported by JSPS KAKENHI Grants Numbers 
JP21H01031 and 19K03752.

\end{acknowledgment}


\begin{thebibliography}{99}
\bibitem{CAMcomment} 
Crystal angular momentum is often called as pseudo-angular momentum as well.
\end{thebibliography}
\end{document}